\documentclass[pdflatex,sn-mathphys-num]{sn-jnl}

\usepackage{graphicx}%
\usepackage{multirow}%
\usepackage{amsmath,amssymb,amsfonts}%
\usepackage{amsthm}%
\usepackage{mathrsfs}%
\usepackage[title]{appendix}%
\usepackage{xcolor}%
\usepackage{textcomp}%
\usepackage{manyfoot}%
\usepackage{booktabs}%
\usepackage{algorithm}%
\usepackage{algorithmicx}%
\usepackage{algpseudocode}%
\usepackage{listings}%

\theoremstyle{thmstyleone}%

\theoremstyle{thmstyletwo}%

\theoremstyle{thmstylethree}%

\begin{document}

\title{Holographic Dark Energy with Torsion}

\author[1]{\fnm{Yongjun} \sur{Yun}}

\author*[1,2]{\fnm{Jungjai} \sur{Lee}}\email{jjlee@daejin.ac.kr}

\affil[1]{\orgdiv{Graduate School Department of Physics}, \orgname{Daejin University}, \orgaddress{\street{1007 Hoguk-ro}, \city{Pocheon}, \postcode{11159}, \state{Gyeonggi}, \country{Korea}}}

\affil[2]{\orgdiv{School of Physics}, \orgname{Korea Institute for Advanced Study}, \orgaddress{\street{85 Hoegi-ro}, \city{Dongdaemun-gu}, \postcode{02455}, \state{Seoul}, \country{Korea}}}

\abstract{We consider the holographic dark energy model with axial torsion which satisfy the cosmological principle. Subsequently, by using the torsional analogues of Friedmann equations for the new equation from Einstein-Cartan gravity theory, we obtain the equation of state for dark energy in this model. We find that the extended holographic dark energy from the particle horizon as the infrared (IR) cut-off does not give the accelerating expansion of the universe. Also, employing the future event horizon as IR cut-off still achieves the accelerating expansion of the universe. In contrast, there is a possibility that the Hubble radius as IR cut-off achieves to the accelerating expansion of the universe in superluminal region for axial torsion. More precisely, the current value of ratio for torsion to the matter density, $\gamma^{0}=0.5$ gives the equation of state of dark energy $\omega_{\Lambda}\cong-1$.}


\maketitle


\section{Introduction}

The accelerating expansion of the universe was discovered in 1998 \cite{R1,R2}. The source of this acceleration, called dark energy having negative pressure and equation of state $\omega_{DE}\cong-1$, still remains an unsolved mystery within the framework of the general relativity (GR). Nevertheless, there is a variety of independent observational evidence supporting the existence of dark energy, such as the type Ia Supernovae (SN Ia) \cite{R1,R3}, the cosmic microwave background (CMB) \cite{R4}, and the baryon acoustic oscillations (BAO) \cite{R5}. The observations have confirmed that the cosmic energy portion of dark energy component consists of roughly 70\% and favors the $\Lambda$CDM model, which is the simplest cosmological model. However, recent improvements in the precision of observations have led to the problem of Hubble tension \cite{R6}, which represents a discrepancy between the Hubble parameters measured from the CMB measurements \cite{R7} and the local distance ladder \cite{R8}. Of course, various dark energy models that replaces the cosmological constant $\Lambda$ have been proposed, however, none of these dark energy models have been proven to be problem-free. Hence, we were motivated to take another new approach, so specifically consider dark energy model with the holographic principle, which can overcome the cosmological constant problem.

Cohen, Kaplan, and Nelson \cite{R9}, inspired by Bekenstein bound \cite{R10}, suggested that an energy bound, which means that in a region of size $L$, the quantum vacuum energy should not exceed the mass of a black hole of the same size. In 2004, based on the energy bound, Miao Li suggested the holographic dark energy (HDE) model and showed the HDE provides the accelerating expansion of the universe, when the future event horizon is chosen as the IR cut-off \cite{R11}. Li's HDE model is in good agreement with the present cosmological observational data \cite{R12}. Therefore, this model has become one of the most competitive candidates for dark energy. But unfortunately, by using the future event horizon, in Li's HDE model and its extensions, there is an inherent problem of causality and circular logic between the definition of future event horizon and the accelerating expansion of the universe and some attempts have been made to solve these problems \cite{R13, R14}. So, we think that it is necessary to investigate HDE in some extended gravity theories if these problems could be resolved. Alternatively, the Einstein-Cartan (EC) gravity theory which can prevent the formation of singularities \cite{R15}, is quite attractive and well-motivated to reconstruct HDE in the extended gravity theory.

The EC theory is a purely geometric extension of GR and was first introduced by Élie Cartan in 1922, who suggested that torsion as the macroscopic manifestation of the intrinsic angular momentum of the matter, namely spin \cite{R16}. Unlike GR which employs both the metric compatibility and the torsion-free condition, the EC theory employs the metric compatibility but not the torsion-free condition. According to this theory, the dynamical variables are the metric tensor and an affine connection. Here, the torsion tensor is mathematically defined as the anti-symmetric part of the affine connection. Therefore, new geometrical degrees of freedom are added. However, a weakness of this theory is that there is still no experimental or observational evidence supporting the existence of torsion in the universe. The main reason is that the torsion effect appears at extremely high densities, such as inside black holes or the early universe \cite{R15}. Despite environmental limitations that make implementation impossible, many experimental tests are underway to detect the effects of torsion at the scale of the solar system \cite{R17}. Among them, there was an attempt to measure torsion through the Lense-Thirring precession effect around the Earth. According to the experiments to detect torsion, the results are still not sufficient to demonstrate an irrefutable discrepancy with the predictions of GR \cite{R17}. Nevertheless, the EC theory is well worth studying. Despite the success of GR, unresolved problems such as the origin of dark energy, dark matter, formation of singularities, and inflation still remain. The EC theory is currently being studied as a promising candidate theory that could solve some of these problems \cite{R18, R19, R20, R21, R22, R23, R24, R25, R26}.

In this paper, we study the HDE model with axial torsion which is totally anti-symmetric one of the non-vanishing components of torsion tensor that satisfy the cosmological principle. The point of emphasis in aspect of the cosmological observations is that the axial torsion preserves the geodesic equation given by the action principle in GR. To obtain the equation of state for dark energy with axial torsion, instead of explicit solution for the scale factor from torsional analogues of Friedmann equations, we directly use differential equations in term of the effective density parameter $\Bar{\Omega}$. Like in the original model, the Hubble radius, the particle horizon, and the future event horizon will be chosen as the IR cut-off and their scaling behaviors will be investigated to get their equation of states. In the original model, the Hubble radius as IR cut-off does not achieve to the accelerating expansion of the universe, because the scaling behavior of matter and dark energy became the same. Finally, by calculating equation of state for dark energy, we will investigate if each IR cut-off achieves the accelerating expansion of the universe in the EC theory.

In Section II, we derive torsional analogues of Friedmann equations from new equation with axial torsion in the EC theory. Subsequently, we calculate equation of state for dark energy and explore the achievement of each IR cut-off to the accelerating expansion of the universe. In Section III,   we will give conclusions and discussions, especially on the possibility of the Hubble radius as IR cut-off. From now on, we will use natural units $c=\hbar=1$, the metric signature $(-,+,+,+)$, the Einstein summation convention, the Greek indices $\mu, \nu, \rho, \cdots$ run from 0
to 3, and the Latin indices $i,j,k, \cdots = 1, 2, 3, $ stand for spatial components.


\section{Extended Holographic Dark Energy}

\subsection{Spacetimes with Torsion}
In the EC theory, the dynamical variables are the metric $\bold{g}$ and the affine connection $\bold{\Gamma}$. Imposing the metric compatibility ${\nabla_\rho}g_{\mu\nu}=0$, where $\nabla_{\rho}$ is the covariant derivative with affine connection, then this connection splits $\Gamma^{\rho}{}_{\mu\nu}=\tilde{\Gamma}^{\rho}{}_{\mu\nu}+K^{\rho}{}_{\mu\nu}$, where $\tilde{\Gamma}^{\rho}{}_{\mu\nu}$ is the Levi-Civita connection, $K^{\rho}{}_{\mu\nu}\equiv S^{\rho}{}_{\mu\nu}+S_{\mu\nu}{}^{\rho}+S_{\nu\mu}{}^{\rho}$ is the contorsion and $S^{\rho}{}_{\mu\nu}\equiv\Gamma^{\rho}{}_{[\mu\nu]}$ is the torsion tensor. 

By using the condition of parallel transport along a curve $x^{\rho}=x^{\rho}(s)$, where $s$ is affine parameter, then we obtain the autoparallel equation:
\begin{equation} \label{Autoparallel equation}
    \frac{d^2{x^{\rho}}}{ds^2} + \Gamma^{\rho}{}_{\mu\nu}\frac{dx^{\mu}}{ds}\frac{dx^{\nu}}{ds} = 0.
\end{equation}
If the torsion is totally anti-symmetric $S_{\rho\mu\nu}=S_{[\rho\mu\nu]}$, then the contorsion becomes $K_{\rho\mu\nu}=S_{\rho\mu\nu}$. Moreover, in the second term of (\ref{Autoparallel equation}), the contorsion contracted part vanishes due to the symmetric property and hence the autoparallel equation coincides with the geodesic equation:
\begin{equation} \label{Geodesic equation}
    \frac{d^2{x^{\rho}}}{ds^2} + \tilde{\Gamma}^{\rho}{}_{\mu\nu}\frac{dx^{\mu}}{ds}\frac{dx^{\nu}}{ds} = 0.
\end{equation}
Obviously, in the Riemann-Cartan spacetime, the totally anti-symmetric components of torsion tensor preserve the geodesic equation given by the action principle in GR and also satisfy the cosmological principle. Based on this fact, to explore dark energy in the EC theory, it seem to be reasonable to use this torsion tensor.

Thus, the 24 independent components of the torsion tensor are to be reduced to 4 components which is known as the axial part of torsion tensor \cite{R27}. Clearly it can be related to an axial vector by using the Levi–Civita tensor $\varepsilon_{\mu\nu\rho\sigma}$ as follows:
\begin{equation} \label{Axial vector of torsion}
    S_{\mu\nu\rho} \equiv \frac{1}{\sqrt{3!}}\varepsilon_{\mu\nu\rho\sigma}N^{\sigma},
\end{equation}
where $N^{\sigma}=(N^{0},\bold{N})$ is the axial vector of the torsion. Here, $N^{0}(t,\bold{x})$ is the time component and $\bold{N}(t,\bold{x})$ is the spatial component of the axial torsion.

Let us consider the action: $S=S_{EC}+S_{M}$, where $S_{M}$ is the action of matter and $S_{EC}$ is the EC action given by
\begin{equation} \label{EC action}
    S_{EC} = \frac{1}{16\pi G}\int d^{4}x\sqrt{-g}R(\bold{g},\bold{\Gamma}).
\end{equation}
In our framework, the Ricci scalar splits into the metric dependent part and the axial torsion part \cite{R28}
\begin{equation} \label{Ricci scalar}
    R(\bold{g},\bold{\Gamma}) = \Tilde{R}(\bold{g}) - N^{\sigma}N_{\sigma},
\end{equation}
where $\Tilde{R}(\bold{g})$ is the usual Ricci scalar with respect to the metric. Subsequently, the variation with respect to the metric gives a new equation:
\begin{equation} \label{New equation}
    \Tilde{R}_{\mu\nu} - \frac{1}{2}g_{\mu\nu}\Tilde{R} - g_{\mu\nu}g_{\rho\sigma}N^{\rho}N^{\sigma} - 2g_{\mu\rho}g_{\nu\sigma}N^{\rho}N^{\sigma} = 8\pi G T_{\mu\nu},
\end{equation}
where we used rescaling $N^{\rho}\rightarrow \sqrt{2}N^{\rho}$. In the new equation, by taking the covariant derivative $\nabla_{\mu}$ with affine connection, and  by using a relation $\nabla_{\mu}A^{\mu\nu}=\Tilde{\nabla}_{\mu}A^{\mu\nu}$ for any symmetric tensor $A^{\mu\nu}$ and the totally anti-symmetric contorsion, we can find
\begin{equation} \label{Energy-momentum}
    \Tilde{\nabla}_{\mu}T^{\mu\nu} = - \frac{1}{8\pi G} \Tilde{\nabla}_{\mu}\left(g^{\mu\nu}g^{\sigma\rho}N_{\sigma}N_{\rho} + 2g^{\mu\rho}g^{\nu\sigma}N_{\sigma}N_{\rho}\right),
\end{equation}
where $\Tilde{\nabla}_{\mu}$ is the usual covariant derivative with Levi-Civita connection. Clearly, the energy-momentum tensor is generally not conserved for the non-zero torsion.

\subsection{FLRW cosmology with Axial Torsion}
Let us expand the FLRW cosmology, which based on the cosmological principle. Specifically, we must check whether the cosmological principle holds in torsional background. It has been known that only torsion tensor such as $S_{[ijk]}$ and $S^{i}{}_{i0}$ satisfies this principle \cite{R29}. Obviously, the homogeneity and isotropy of the universe is preserved when the axial torsion (\ref{Axial vector of torsion}) takes the form $N^{0}(t)\neq0$ and $\bold{N}=0$.

Specifically, according to Milton's work \cite{R30}, let us define the causality regions of torsion: subluminal, superluminal and luminal at any point in spacetime. The subluminal region $k\equiv g_{\mu\nu}N^{\mu}N^{\nu}=-(N^{0})^{2}+N^{2}<0$, it is possible to move to the frame such that $N^{0}(t,\bold{x})\neq0$ and $\bold{N}=0$. If we put $N^{0}=N^{0}(t)$, there are the homogeneity and isotropy in this frame. The superluminal region $k>0$, it is impossible to move to the frame such that $\bold{N}=0$, thus the cosmological principle is not guaranteed. The same goes for the luminal region $k=0$.

However, the current cosmological observations are still not sufficient to demonstrate an irrefutable discrepancy with the predictions of GR \cite{R17}. Nevertheless, for the EC theory to be viable, fluctuations in the torsion field should be truncated at perhaps at the Planck length scale \cite{R30}. This constraint blocks the propagation of torsion. Therefore, it may be interpreted that the effect of torsion is difficult to detect in the universe around us. Based on this constraint, although unsatisfactory, we require that the cosmological principles are maintained approximately despite the presence of non-zero $\bold{N}$ at macroscopic scale.

In EC theory, treating independently the metric $\bold{g}$ and the connection $\Gamma$ means that the line element of the spacetime is identical to its GR counterpart. Hence, we can employ the FLRW metric of GR to that of EC. Subsequently, the new equation (\ref{New equation}) for the flat FLRW metric turn into the torsional analogues of Friedmann equations:
\begin{equation} \label{Torsional analogues of Friedmann equations}
    H^{2} = \frac{1}{3M_{P}^{2}}\rho - \frac{1}{3}\left(N^{2}-3\left(N^{0}\right)^{2}\right), \quad
    \Dot{H} + H^{2} = -\frac{1}{6M_{P}^{2}}\left(\rho+3p\right) - \frac{4}{3}N^{2},
\end{equation}
where $N^{2}\equiv g_{ij}N^{i}N^{j}$. Here, $\rho=\rho_{m}+\rho_{\Lambda}$ and $p=p_{m}+p_{\Lambda}$ are the sum of matter energy density and vacuum energy density and the sum of corresponding pressure components respectively. These equations (\ref{Torsional analogues of Friedmann equations}) can be rewritten as equivalent forms:
\begin{equation} \label{Equivalent form}
     H^{2} = \frac{1}{3M_{P}^{2}}\Bar{\rho}, \quad
     2\Dot{H} + 3H^{2} = -\frac{1}{M_{P}^{2}}\Bar{p}.
\end{equation}
Here, $\Bar{\rho}\equiv\rho_{m}+\rho_{\Lambda}+\rho_{T}$ is effective energy density and $\Bar{p}\equiv p_{m}+p_{\Lambda}+p_{T}$ is effective pressure, where $\rho_{T} \equiv-M_{P}^{2}(N^{2}-3(N^{0})^{2})$ is not an energy density and $p_{T}\equiv M_{P}^{2}(3N^{2}-(N^{0})^{2})$ is not a pressure. Note that both $\rho_{T}$ and $p_{T}$ are geometric quantities and $\Bar{\rho}$ is always positive. Similarly, we define effective critical density $\Bar{\rho}_{c}\equiv 3M_{P}^{2}H^{2}$ and effective density parameter $\Bar{\Omega}\equiv \rho/\Bar{\rho}_{c}$, then, from equation(\ref{Equivalent form}), we get
\begin{equation} \label{First eq.}
    1 = \Bar{\Omega}_{m} + \Bar{\Omega}_{\Lambda} + \Bar{\Omega}_{T}.
\end{equation}
By using these equations (\ref{Torsional analogues of Friedmann equations}), we obtain effective continuity equation:
\begin{equation} \label{Effective continuity equation}
    \Dot{\Bar{\rho}} + 3H\left(\Bar{\rho}+\Bar{p}\right) = 0,
\end{equation}

As discussed before \cite{R17}, because the observational torsion effect up to now is extremely tiny, we wish to align the analysis with assumptions of standard cosmology as far as possible, we require the conservation of energy-momentum tensor $\Tilde{\nabla}_{\mu}T^{\mu\nu}=0$ for non-interacting individual energy density species \cite{R31}. From this, we could use the continuity equation for matter and vacuum energy. Eventually, from the equation (\ref{Effective continuity equation}) we get
\begin{equation} \label{Torsion term}
    \Dot{\rho}_{T} + 3H\left(\rho_{T}+p_{T}\right)=0,
\end{equation}
where (\ref{Torsion term}) is not a genuine continuity equation but torsional analogue of continuity equation. Furthermore, we find the scaling behaviors $\rho_{m}=\rho^{0}_{m}a^{-3}$, $\rho_{\Lambda}=\rho^{0}_{\Lambda}a^{-3(1+\omega_{\Lambda})}$ and $\lvert\rho_{T}\lvert=\lvert\rho^{0}_{T}\lvert a^{-3(1+\omega_{T})}$, where the superscript $0$ represents the present, $\omega_{m}$ is the equation of state for matter and $\omega_{\Lambda}$ is the equation of state for vacuum energy. Here,
\begin{equation} \label{Omega T}
    \omega_{T} \equiv \frac{p_{T}}{\rho_{T}} = -\frac{3N^{2}-\left(N^{0}\right)^{2}}{N^{2}-3\left(N^{0}\right)^{2}},
\end{equation}
which is not an equation of state. Consequently, we preserved the scaling behavior of standard cosmology and built a torsional analogue of scaling behavior.

\subsection{Holographic Dark Energy Model with torsion}
Cohen, Kaplan, and Nelson \cite{R9}, inspired by the Bekenstein bound, suggested that in a region of size $L$, the quantum vacuum energy $L^{3}\Lambda^{4}$ should not exceed the mass of a black hole of the same size, $LM_{P}^{2}$, where $L$ and $\Lambda$ are infrared (IR) and ultraviolet (UV) cut-off, respectively. The vacuum energy density $\rho_{\Lambda}\sim\Lambda^{4}$, thus $L^{3}\rho_{\Lambda}\lesssim LM_{P}^{2}$. Saturating in this bound: 
\begin{equation} \label{HDE density}
    \rho_{\Lambda} = 3c^{2}M_{P}^{2}L^{-2},
\end{equation}
where $\rho_{\Lambda}$ called dark energy density and $c$ is a free parameter. 

To we obtain the current equation of state for dark energy, taking the Taylor expansion
\begin{equation} \label{Expanding}
    \ln{\rho_{\Lambda}} = \ln{\rho^{0}_{\Lambda}} + \frac{d\ln{\rho_{\Lambda}}}{d\ln{a}}\Big{|}_{0}\ln{a} + \frac{1}{2!}\frac{d^{2}\ln{\rho_{\Lambda}}}{d(\ln{a})^{2}}\Big{|}_{0}(\ln{a})^{2} + \cdots,
\end{equation}
where the derivatives are taken at the present $a_{0} = 1$. By putting the scaling behavior $\rho_{\Lambda}=\rho^{0}_{\Lambda}a^{-3(1+\omega_{\Lambda})}$ into (\ref{Expanding}), then
\begin{equation} \label{EoS HDE}
    \omega_{\Lambda} \sim - 1 - \frac{1}{3}\frac{d\ln{\rho_{\Lambda}}}{d\ln{a}}\Big{|}_{0},
\end{equation}
up to the first order. By using the torsional analogue of Friedman equation (\ref{First eq.}), we obtain (see appendix A)
\begin{equation} \label{dark energy density}
    \rho_{\Lambda} 
    \sim \frac{\Bar{\Omega}_{\Lambda}}{1-\Bar{\Omega}_{\Lambda}}a^{3}\left(1+\gamma^{0}a^{-3\omega_{T}}\right),
\end{equation}
and
\begin{equation} \label{First derivative}
    \frac{d\ln{\rho}_{\Lambda}}{d\ln{a}} \sim -3 + \frac{\Bar{\Omega}_{\Lambda}^{'}}{\Bar{\Omega}_{\Lambda}\left(1-\Bar{\Omega}_{\Lambda}\right)} - 3\omega_{T}\frac{\gamma^{0}a^{-3\omega_{T}}}{1+\gamma^{0}a^{-3\omega_{T}}},
\end{equation}
where $\gamma^{0}\equiv\lvert\rho^{0}_{T}\lvert/\rho^{0}_{m}$ represents the ratio between the torsion and the energy density of the matter at the present time, and will have to be determined through experimentation and observation.

In the present work, we take $L$ as the size of the current universe, for three typical cases, (I) the Hubble radius, (II) the particle horizon and (III) the future event horizon. (I) The Hubble radius is chosen as the IR cut-off $L=H^{-1}$. By definition (\ref{HDE density}), the dark energy density becomes $\rho_{\Lambda} = 3c^{2}M_{P}^{2}H^{2}$. Using this density and the effective density parameter $\Bar{\Omega}_{\Lambda}=\rho_{\Lambda}/\Bar{\rho}_{c}$, then we get $\Bar{\Omega}_{\Lambda}=c^{2}$, thus its first derivative:
\begin{equation} \label{Differential Omega in Hubble}
    \Bar{\Omega}_{\Lambda}^{'}=0,
\end{equation}
where $'\equiv d/d\ln{a}$. Putting (\ref{Differential Omega in Hubble}) into (\ref{First derivative}), and using (\ref{EoS HDE}), we arrive at
\begin{equation} \label{EoS HDE in Hubble radius}
    \omega_{\Lambda} 
    \sim \omega_{T}\frac{\gamma^{0}}{1+\gamma^{0}},
\end{equation}
where $\omega_{\Lambda}$ does not depend on a free parameter $c$, and $\gamma^{0}$ is always positive. In the subluminal where $\bold{N}=0$, by using (\ref{Omega T}), we get $\omega_{T}=-1/3$, thus $\omega_{\Lambda}>-1/3$. In the luminal, $\omega_{T}=1$, thus $0<\omega_{\Lambda}<1$. Obviously, there is no accelerating expansion of the universe. On the other hand, in the superluminal where $N^{0}=0$, we get $\omega_{T}=-3$, thus $\omega_{\Lambda}<-3$. Interestingly, there is accelerating expansion of the universe. 

Moreover, using the redshift parameter $z$ related to $a=1/(1+z)$, we find $\gamma(z_{CMB}=1100)\sim10^{-28}$ at the CMB formation. Consequently, the contribution of torsion could be neglected. According to current cosmological observations, the $\rho_{\Lambda}$ called dark energy density for dark energy has $\omega_{DE}\cong-1$. If $\gamma^{0}\cong0.5$, then $\omega_{\Lambda}\cong-1$, which means the solution exists that explains the current observation. In the future for very large $a$, then the equation of state for dark energy becomes asymptotically $\omega_{\Lambda}\sim-3$. Eventually, the accelerating expansion of the universe increases monotonically up to $\omega_{\Lambda}\sim-3$.

(II) The particle horizon is chosen as the IR cut-off $L=R_{H}$, which is given by
\begin{equation} \label{Particle horizon}
    R_{H} \equiv a\int_{0}^{t}\frac{dt}{a} = a\int_{0}^{a}\frac{da}{Ha^{2}}.
\end{equation}
By using dark energy density becomes $\rho_{\Lambda}=3c^{2}M^{2}_{p}R_{H}^{-2}$ and the effective density parameter $\Bar{\Omega}_{\Lambda}=\rho_{\Lambda}/\Bar{\rho}_{c}$, we get
\begin{equation} \label{RH}
    R_{H} 
    = \frac{c}{H\sqrt{\Bar{\Omega}_{\Lambda}}},
\end{equation}
where $c$ is always positive in the expanding universe. Putting (\ref{RH}) into (\ref{Particle horizon}),
\begin{equation} \label{RH-1}
    \int_{0}^{a}\frac{da}{Ha^{2}} = \frac{c}{\sqrt{\Bar{\Omega}_{\Lambda}}}\frac{1}{Ha},
\end{equation}
and taking the derivative with respect to $'\equiv d/d\ln{a}$ on both sides of (\ref{RH-1}), we find the first torsional analogue of Friedmann equation with the particle horizon as IR cut-off:
\begin{equation} \label{Differential equation in Particle horizon}
    \frac{\Bar{\Omega}_{\Lambda}^{'}}{\Bar{\Omega}_{\Lambda}\left(1-\Bar{\Omega}_{\Lambda}\right)} 
    = 1 - \frac{2\sqrt{\Bar{\Omega}_{\Lambda}}}{c} + 3\omega_{T}\frac{\gamma^{0}a^{-3\omega_{T}}}{1+\gamma^{0}a^{-3\omega_{T}}}.
\end{equation}
Putting (\ref{Differential equation in Particle horizon}) into (\ref{First derivative}), and using (\ref{EoS HDE}), we arrive at arrive at
\begin{equation} \label{EoS HDE in particle horizon}
    \omega_{\Lambda} \sim - \frac{1}{3} + \frac{2}{3c}\sqrt{\Bar{\Omega}^{0}_{\Lambda}}, \quad 
    \omega_{\Lambda} > - \frac{1}{3}.
\end{equation}
Explicitly, there is no accelerating expansion of the universe in all causality regions of the torsion.

(III) The future event horizon is chosen as the IR cut-off $L=R_{h}$, which is given by
\begin{equation} \label{Future event horizon}
    R_{h} \equiv a\int_{t}^{\infty}\frac{dt}{a} = a\int_{a}^{\infty}\frac{da}{Ha^{2}}.
\end{equation}
Compared to the particle horizon, only the sign of the $c$ term is different. Consequently, we arrive at
\begin{equation} \label{Eos HDE in future event horizon}
    \omega_{\Lambda} \sim - \frac{1}{3} - \frac{2}{3c}\sqrt{\Bar{\Omega}^{0}_{\Lambda}}, \quad
    \omega_{\Lambda} < - \frac{1}{3}.
\end{equation}
Clearly, there is accelerating expansion of the universe. In the subluminal where $\bold{N}=0$ and the luminal, then $\rho_{T}>0$, thus $\Bar{\Omega}_{\Lambda}<\Omega_{\Lambda}$. Consequently, the torsion acts to decrease the magnitude of accelerating expansion. On the other hand, in the superluminal where $N^{0}=0$, then $\rho_{T}<0$, thus $\Bar{\Omega}_{\Lambda}>\Omega_{\Lambda}$. In contrast to the two regions, this magnitude increases.


\section{CONCLUSIONS}

In this work, assuming the axial torsion which is totally anti-symmetric nonvanishing components of torsion tensor, we derived the new field equation from the Einstein-Cartan action, and showed that the energy-momentum tensor is generally not conserved. Because the axial torsion satisfys the cosmological principle and also preserves the geodesic equation of GR, we think that the assumptions are very reasonable. Totally anti-symmetrice torsion tensor could be re-expressed as the axial vectors $N^{\mu}$. Subsequently, by using the flat FLRW metric, we obtained the torsional analogues of Friedmann equations from the new field equation. Because  to date there have been no experimental or observational reports that deviate from the predictions of GR, we wished to align the analysis with assuptions of GR as far as possible. Thus, we required conservation of the energy-momentum tensor for individual energy density species. This requirement allowed us to employ the scaling behaviors of the standard cosmology that align well with current observations. Consequently, in our framework, the torsion effect is linked only to the torsional analogues of Friedman equations.


The main results of the HDE model with torsion can be summarized: First, when the particle horizon is chosen as the IR cut-off, then $\omega_{\Lambda}>-1/3$ like HDE model without torsion. Thus we got the result that it could not achieve the accelerating expansion of the universe. Second, in case of the future event horizon as the IR cut-off, then $\omega_{\Lambda}<-1/3$ like HDE model without torsion, there is still the accelerating expansion of the universe. In this case, we got the torsional effect diminishing the accelerating expansion in the subluminal and luminal regions. On the other hand, in the superluminal region, the torsion acts to raise the accelerating expansion. 

Interestingly, third, when the Hubble radius is chosen as the IR cut-off, then the equation of state of dark energy is not zero unlike the torsion-free case, which is independent of the free parameter $c$. Especially, in the superluminal region, then $\omega_{\Lambda}<-3$ with $\omega_{T}=-3$, thus the HDE with torsion could provide accelerating expansion of the universe. On the other hand, the subluminal and luminal region are not. Therefore, in contrast to the model without torsion, the Hubble radius as the IR cut-off may achieve the accelerating expansion, so is likely to be a candidate for dark energy. In the paste, when the CMB was formed, $\gamma_{CMB}\sim10^{-28}$, so the contribution of torsion could be ignored. According to current observation, the equation of state for dark energy has $\omega_{DE}\cong-1$. If $\gamma^{0}\cong0.5$, then $\omega_{\Lambda}\cong-1$, which means there exists a solution that explains the current observation. Of course, $\gamma^{0}$ must be determined through experiments measuring the effect of torsion. In the future, for large $a$, then $\gamma\sim a^{9}$, so the contribution of torsion increases monotonically up to $\omega_{\Lambda}\sim-3$ from the current value $\omega_{\Lambda}\cong-1$. Therefore in the Holographic dark model with torsion, not only the future event horizon but also the Hubble radius could be chosen as the IR cut-off and its vacuum energy could be a candidate for dark energy.

The inclusion of the Hubble radius may provide significant advantages which is to avoid the causality  and circular logic problems associated with the future event horizon in the holographic dark energy model.


\bmhead{Acknowledgements}

This work was supported by the Daejin University Research Grants in 2024.

\begin{appendices}

\section{}
\subsection{First derivative}
Using effective critical density $\Bar{\rho}_{c}\equiv 3M_{P}^{2}H^{2}$ and effective density parameter for dark energy $\Bar{\Omega}_{\Lambda} \equiv \rho_{\Lambda}/\Bar{\rho}_{c}$, we get
\begin{equation} \label{A.1}
    \rho_{\Lambda} 
    = \Bar{\rho}_{c} \Bar{\Omega}_{\Lambda} = 3M^{2}_{P}H^{2}\Bar{\Omega}_{\Lambda} \sim H^{2} \Bar{\Omega}_{\Lambda}.
\end{equation}
From the first torsional analogue of Friedmann equation (\ref{First eq.}),
\begin{equation} \label{A.2}
     1 - \Bar{\Omega}_{\Lambda} = \Bar{\Omega}_{m} + \Bar{\Omega}_{T} = \frac{\rho_{m}}{\Bar{\rho}_{c}} + \frac{\rho_{T}}{\Bar{\rho}_{c}},
\end{equation}
and by putting the scaling behaviors $\rho_{m}=\rho^{0}_{m}a^{-3}$ and $\lvert\rho_{T}\lvert=\lvert\rho^{0}_{T}\lvert a^{-3(1+\omega_{T})}$ into (\ref{A.2}),
\begin{equation} \label{A.3}
     1 - \Bar{\Omega}_{\Lambda} =  \Bar{\Omega}^{0}_{m} H^{2}_{0}H^{-2}a^{-3} \left(1+\gamma^{0}a^{-3\omega_{T}}\right) \sim H^{-2}a^{-3}\left(1+\gamma^{0}a^{-3\omega_{T}}\right),
\end{equation}
where $\gamma^{0}\equiv\lvert\rho^{0}_{T}\lvert/\rho^{0}_{m}$ represents the ratio between the torsion and the energy density of the matter at the present time. Putting (\ref{A.1}) into (\ref{A.3}), then dark energy density:
\begin{equation} \label{A.4}
    \rho_{\Lambda} 
    \sim \frac{\Bar{\Omega}_{\Lambda}}{1-\Bar{\Omega}_{\Lambda}}a^{-3}\left(1+\gamma^{0}a^{-3\omega_{T}}\right).
\end{equation}
Using (\ref{A.4}), we obtain the first derivative:
\begin{equation} \label{A.5}
    \frac{d\ln{\rho}_{\Lambda}}{d\ln{a}} \sim -3 + \frac{\Bar{\Omega}_{\Lambda}^{'}}{\Bar{\Omega}_{\Lambda}\left(1-\Bar{\Omega}_{\Lambda}\right)} - 3\omega_{T}\frac{\gamma^{0}a^{-3\omega_{T}}}{1+\gamma^{0}a^{-3\omega_{T}}},
\end{equation}
where $'\equiv d/d\ln{a}$.

\end{appendices}


\end{document}